\documentclass[11pt,twoside]{article}
\usepackage{./asp2014}

\aspSuppressVolSlug
\resetcounters

\begin{document}

\title{Similarities in the structure of the circumstellar environments of B[e] 
supergiants and yellow hypergiants}
\author{Anna Aret,$^1$ Indrek Kolka,$^1$ Michaela Kraus,$^{1,2}$ and Grigoris Maravelias$^2$
\affil{$^1$Tartu Observatory, 61602 T{\~o}ravere, Tartumaa, Estonia; \email{aret@to.ee}\\
$^2$Astronomick\'y \'ustav AV\v{C}R, v.v.i., Ond\v{r}ejov, Czech Republic}}

\paperauthor{Anna Aret}{aret@to.ee}{}{Tartu Observatory}{}{T{\~o}ravere}{Tartumaa}{61602}{Estonia}
\paperauthor{Indrek Kolka}{indrek@to.ee}{}{Tartu Observatory}{}{T{\~o}ravere}{Tartumaa}{61602}{Estonia}
\paperauthor{Michaela Kraus}{kraus@to.ee}{}{Tartu Observatory}{}{T{\~o}ravere}{Tartumaa}{61602}{Estonia}
\paperauthor{Grigoris Maravelias}{maravelias@asu.cas.cz}{}{Astronomick\'y \'ustav AV\v{C}R}{}{Ond\v{r}ejov}{}{25165}{Czech Republic}

\begin{abstract}
Yellow Hypergiants (YHGs) and B[e] supergiants (B[e]SGs), though in different 
phases in their evolution, display many features in common. This is partly due
to the fact that both types of objects undergo strong, often asymmetric mass
loss, and the ejected material accumulates in shells, rings, or disk-like 
structures, giving rise to emission from warm molecules and dust. 
We performed an optical spectroscopic survey of northern 
Galactic emission-line stars aimed at identifying tracers for the 
structure and kinematics of circumstellar environments. We identified
two sets of lines, [O\,{\sc i}] and [Ca\,{\sc ii}], which originate from
the discs of B[e]SGs. The same set of lines is observed in V1302 Aql and 
V509 Cas, which are both hot YHGs. While V1302\,Aql is known to have a disc-like 
structure, the kinematical broadening of the lines in V509\,Cas suggest a 
Keplerian disk or ring around this star alike their hotter B[e]SG
counterparts.  
\end{abstract}

\section{Introduction}\label{introduction}
Despite their different evolutionary phases, B[e] supergiants (B[e]SGs) and 
yellow hypergiants (YHGs) share a number of common properties regarding their 
circumstellar environments. Both types of stars experience phases of strongly 
enhanced mass-loss, and the released material accumulates in (multiple) shells, 
bipolar nebulae, and/or disc-like structures, often veiling the central object. 
Moreover, the physical conditions in the envelopes of these stars are ideal 
for molecule and dust condensation. Warm molecular gas is obvious from CO band 
emission \citep*[e.g.,][]{1988ApJ...334..639M, 2006ApJ...651.1130G, 
2013A&A...558A..17O}, and the dust is traced by its infrared excess emission
\citep{1986A&A...163..119Z, 1996MNRAS.280.1062O}. While the enhanced 
mass-loss and eruptions in YHGs are probably caused by an increased pulsation 
activity, the physical mechanism leading to the formation of the dense winds 
and Keplerian discs and rings observed in B[e]SGs is yet unknown, although
pulsations might play a role as well \citep{2016A&A...Kraus}. 

The evolutionary state of B[e]SGs is also still unclear. Based on the observed 
enrichment in $^{13}$C in their discs, \citet{2013A&A...558A..17O} proposed 
that B[e]SGs have just evolved off the main sequence. On the other hand,
their dense dusty environments seem to speak in favour of a post-red supergiant
(post-RSG) evolutionary phase. YHGs 
may have passed through the RSG phase and evolve back to the 
blue, hot side of the Hertzsprung-Russell diagram (HRD). On their journey, they 
encounter the Yellow Void region, in which their envelopes become unstable 
and are successively ejected during a series 
of outbursts \citep[e.g.,][]{2003ApJ...583..923L, 2009ASPC..412...17O}. 
While the progenitor stars of B[e]SGs spread over the full mass range of 
massive stars ($M>8$\,M$_{\odot}$), YHGs are only found in the mass range 
25\,M$_{\odot}<M<50$\,M$_{\odot}$. Nevertheless, it has been suggested that 
YHGs may be evolving toward the B[e] supergiant phase 
\citep*{2007ApJ...671.2059D}. Such a possible evolutionary link in this 
specific mass range should be investigated.

\section{Observations}\label{observations}

During the years 2010--2015, we carried out an optical spectroscopic survey of 
a large sample of Ga\-lac\-tic northern emission-line stars in diverse 
evolutionary states (pre-main sequence stars, classical Be stars, B[e] stars 
including two B[e]SGs, compact planetary nebulae, and YHGs). 
This survey was aimed at identifying characteristic emission features that help 
to study the structure and kinematics of dense circumstellar environments. 
Motivated by the results from previous studies \citep*{2007A&A...463..627K,2010A&A...517A..30K,2012MNRAS.423..284A}, we focused on the strategic 
forbidden emission lines of [O\,{\sc i}] and [Ca\,{\sc ii}], whose
appearance requires high-density environments combined with large
emitting volumes, conditions which are typically met in geometrically thick and
dense circumstellar discs \citep*{2016MNRAS.456.1424A}.

The observations were obtained using the Coud\'{e} spectrograph attached to the 
Perek 2-m telescope at Ond\v{r}ejov Observatory \citep{2002PAICz..90....1S}.
Spectra were taken in three different wavelength regions:
around H~$\alpha$ (6250--6760\,\AA, $R\simeq$ 13\,000),
in the region of the [Ca\,{\sc ii}] $\lambda\lambda$7291,~7324 lines
(6990--7500\,\AA, $R\simeq$ 15\,000), and in the region of the
Ca\,{\sc ii} IR triplet (8470--8980\,\AA, $R\simeq$ 18\,000).
The H~$\alpha$ region also encloses the two [O\,{\sc i}]
$\lambda\lambda$6300, 6364 lines.

\begin{table}[!ht]
\caption{Identification of disc tracers in B[e]SGs (top) and YHGs (bottom).}
\smallskip
\begin{center}
{\small
\begin{tabular}{lllcccc}
\tableline
\noalign{\smallskip}
Object      &           & Sp.type     & Class & [O\,{\sc i}]& [Ca\,{\sc ii}] &Ca\,{\sc ii} IR\\
\noalign{\smallskip}
\tableline
\noalign{\smallskip}
V1478 Cyg  &MWC 349A     & B0-B1.5\,I & B[e]SG &+& +  &emis     \\[3pt]
3 Pup      &HD 62623     & A2.7\,Ib   & A[e]SG &+& +  &emis      \\[5pt]
\noalign{\smallskip}
\tableline
\noalign{\smallskip}
V1302 Aql  & IRC +10420     & A2-F8Ia  & hot YHG   &+& +   &emis      \\[3pt]
V509 Cas   & HR 8752        & A7-G5Ia  & hot YHG   &+& +   &emis/abs  \\[3pt]
$\rho$ Cas & HD 224014      & F0-G7Ia  & cool YHG  &-- &+  &abs         \\[3pt]
V1427 Aql  &HD 179821      & F3-G5Ia   & cool YHG  & --& + &abs      \\[5pt]
\noalign{\smallskip}
\tableline
\end{tabular}
}
\end{center}
\label{tab:SGvsYHG}
\end{table}

The survey contains four YHGs, which we monitor to study variations
in their environments and to trace their outburst activity during their journey
through the Yellow Void instability region. These are discussed in the
following in comparison to the two B[e]SGs (Table\,\ref{tab:SGvsYHG}), which
were presented and discussed in detail in \citet{2016MNRAS.456.1424A}.
The objects are listed in Table\,\ref{tab:SGvsYHG} together with a brief
summary of the important spectral features. Portions of the optical spectra
of the YHGs covering the strategic lines are shown in Fig.\,\ref{fig1} in
comparison to the spectrum of the A[e]SG star 3\,Pup.

\articlefigure{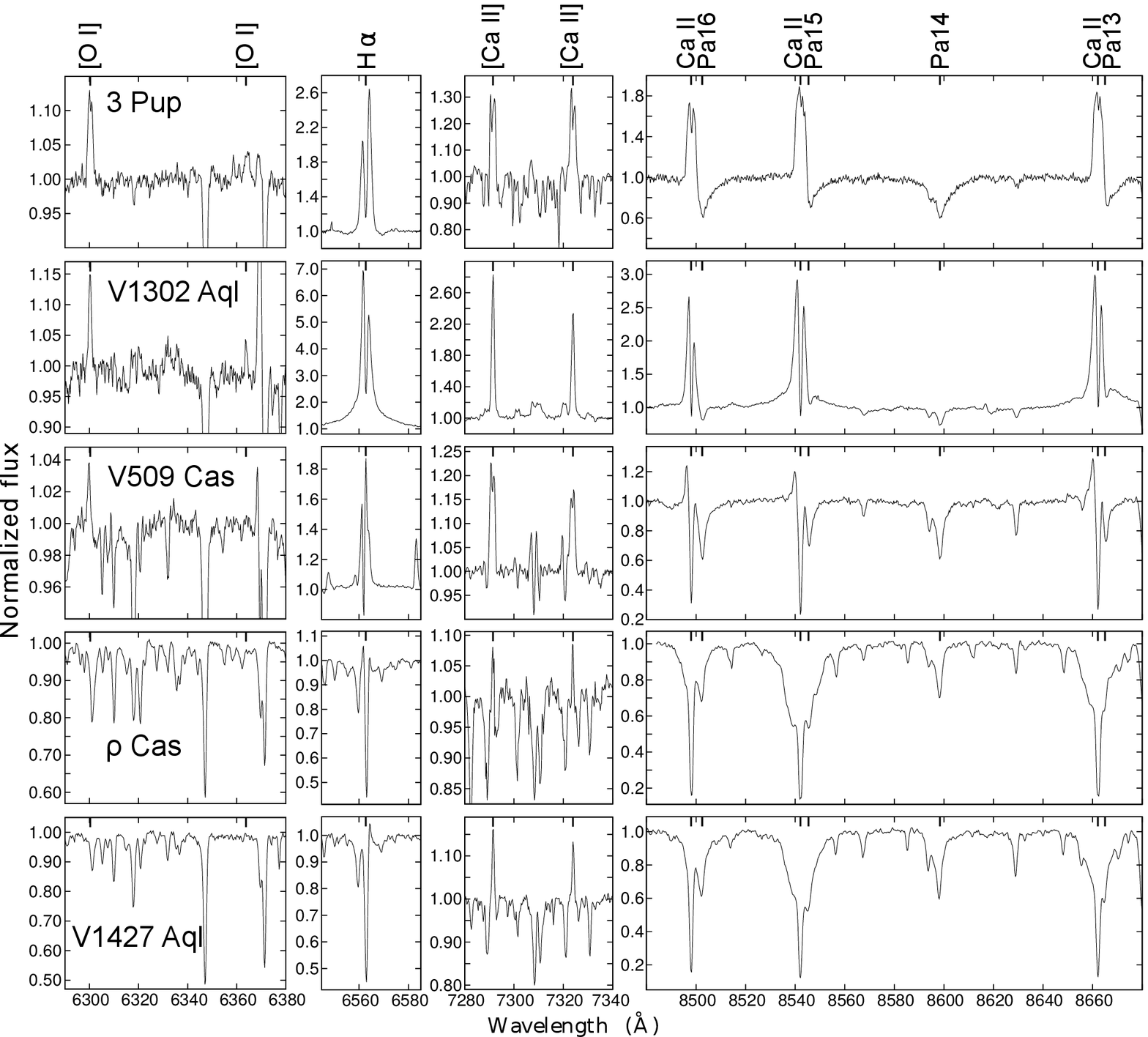}{fig1}{Spectra of the B[e]SG 3\,Pup
\textit{vs} the hot YHGs V1302\,Aql \& V509\,Cas and the cool YHGs $\rho$\,Cas \&
V1427\,Aql.}

\section{Results}\label{results}
Our YHG sample splits into two categories, hot and cool. We define hot YHGs as
those with spectral types A outside outburst. Two of our stars, V1302\,Aql and
V509\,Cas, fall into this group, while the other two objects, $\rho$\,Cas and
V1427\,Aql, belong to the group of cool YHGs. These are characterized by
spectral types F outside outburst.

We identified the [Ca\,{\sc ii}] lines in all stars of our YHG sample, but the
[O\,{\sc i}] lines only in the hot YHGs. In addition, only the hot YHGs display
the Ca\,{\sc ii} triplet lines in emission. To understand this behaviour, we 
recall that the upper levels, from which the [Ca\,{\sc ii}] lines emerge, are 
the lower levels to which the triplet lines decay. In addition, the upper 
levels of the forbidden lines can be populated collisionally from the ground 
level. The fact that we observe the triplet lines in emission only in the hot 
YHGs indicates that the excitation mechanism of the [Ca\,{\sc ii}] lines is 
different in the two groups of stars, meaning that in the cool YHGs only 
collisional level excitation is at work. Moreover, the low effective 
temperature of the cool YHGs translates into a cooler temperature of the stars' 
environment, which suppresses effective collisional population of the levels 
of both Ca\,{\sc ii} and O\,{\sc i}. This explains why the [Ca\,{\sc ii}] lines 
are much weaker and the [O\,{\sc i}] lines are even absent in the spectra of 
these stars. 

\articlefigure{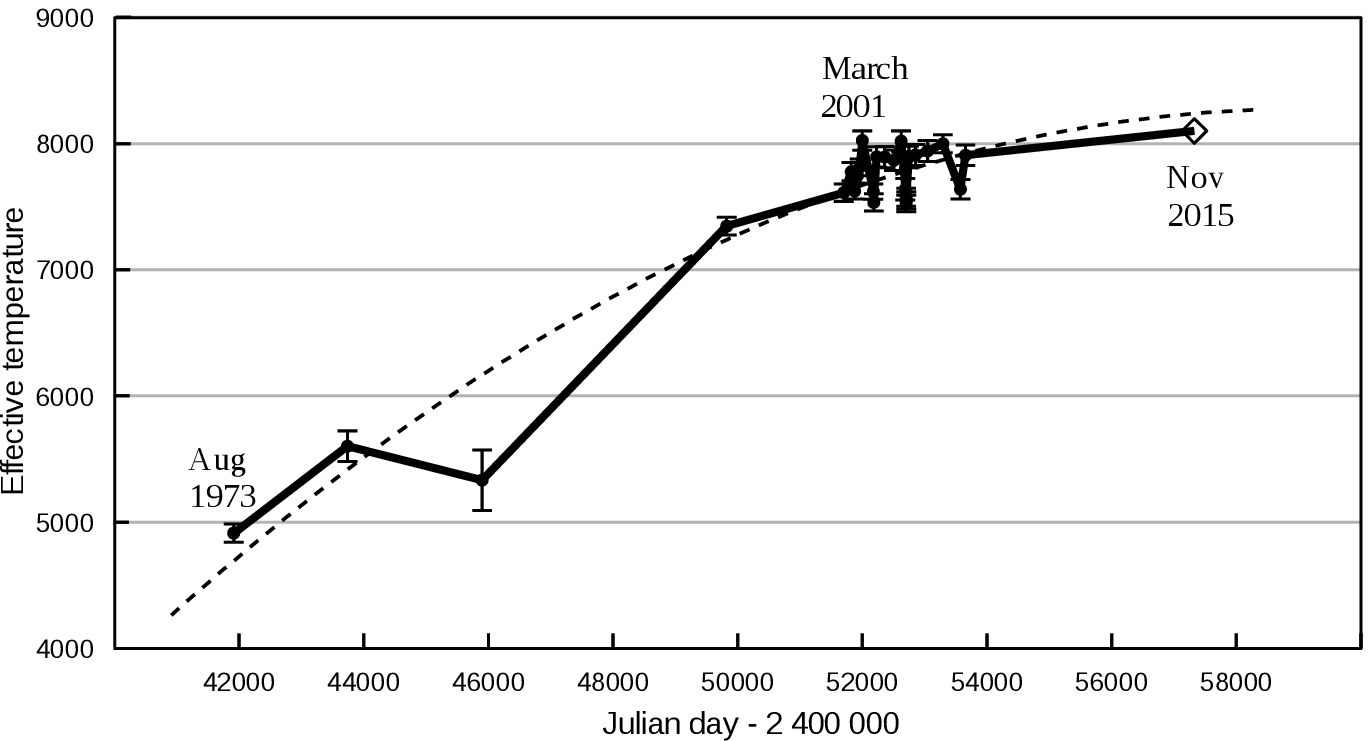}{fig2}{Bold line marks the temperatures of
V509\,Cas from \citet{2012A&A...546A.105N} with one point of November 2015
from our spectra. Added is a dashed polynomial trend line.}

$\rho$\,Cas is famous for its outbursts (or shell episodes), during which the 
star ejects large amounts of material \citep[][Aret et al., this 
volume]{2003ApJ...583..923L}, however, no large-scale nebulosity has been 
detected \citep*{2006AJ....131..603S}. The situation is different for V1427\,Aql,
which has a clearly detached large-scale dust shell \citep{2007A&A...465..457C}.
Not much is known about the shape of the environments of these two objects on 
small scales. The profiles of their [Ca\,{\sc ii}] lines are single-peaked in 
both cool YHGs (Fig.\,\ref{fig1}). Hence, based on our data we have no 
information in the kinematics in their [Ca\,{\sc ii}] line-forming regions. 

The presence of both sets of forbidden lines in the two hot YHGs indicates that 
the physical conditions in their environments could be similar to those in the 
B[e]SGs. The kinematics obtained from the [O\,{\sc i}] and [Ca\,{\sc ii}] line 
profiles of the B[e]SG stars V1478\,Cyg and 3\,Pup agrees with an origin of the 
lines in Keplerian rotating rings \citep{2016MNRAS.456.1424A}. Thus, the same 
scenario might also hold for the hot YHGs. 

The object V1302\,Aql is surrounded by a large-scale spherical shell, a possible 
remnant from the previous RSG state. It is famous for its extreme infrared 
excess indicating large amounts of warm dust, and it loses mass at very high 
rate in an axi\-symmetric wind \citep{2007ApJ...671.2059D}. Moreover, a 
disk-like structure, embedded in the innermost shell, was proposed by 
\citet{2007A&A...465..457C}, and an orientation of the star close to pole-on 
was suggested by \citet{2010AJ....140..339T}. Inspection of its forbidden 
emission lines (Fig.\,\ref{fig1}) reveals that all profiles are very narrow and 
single peaked. Despite the lack of kinematical information, such profiles are 
in agreement with their formation in the nearly pole-on seen disk.

\articlefigure{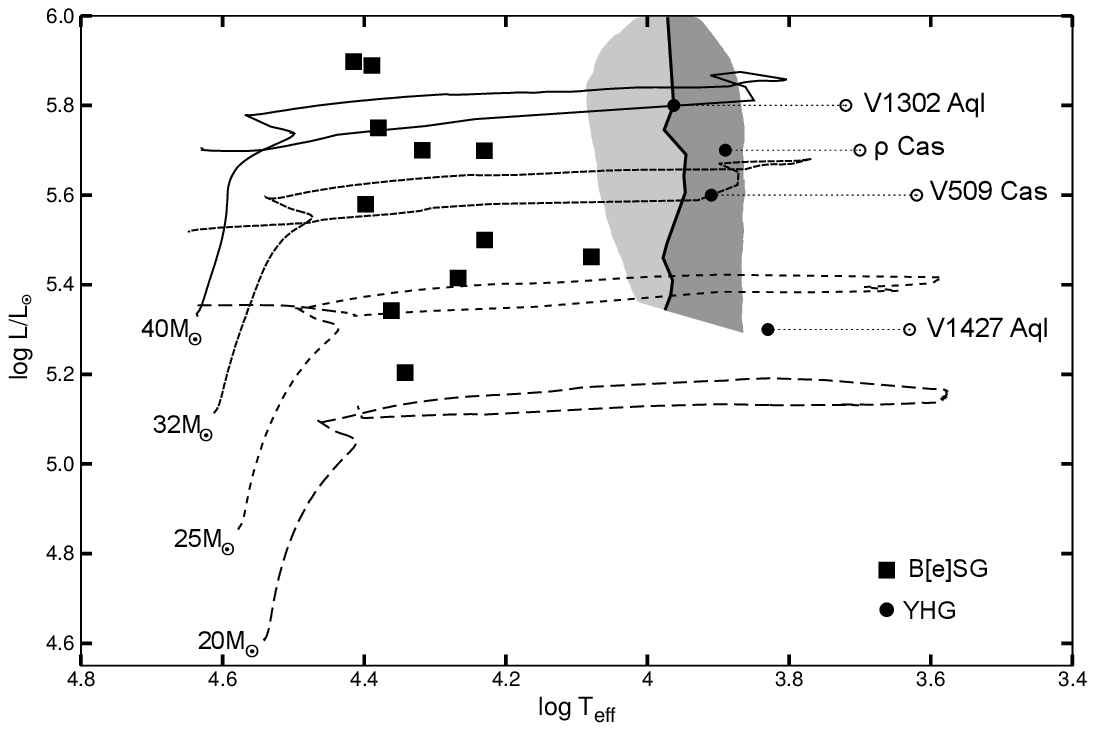}{fig3}{Upper part of the HRD
illustrates possible evolutionary link between YHGs (circles) and
B[e]SGs (squares). For the hypergiants horizontal excursions due to
$T_{\rm eff}$ variations are shown. Evolutionary tracks from
\citet{2012A&A...537A.146E}. Yellow Void is drawn as in
\citet{1997MNRAS.290L..50D}, "first" instability region is darker and it's
high-temperature boundary is marked by solid line \citep{2012A&A...546A.105N}.}

Alike $\rho$\,Cas, V509\,Cas displays no evidence for large-scale shells or 
ejecta \citep{2006AJ....131..603S}, indicating that no high mass loss events 
took place prior to 500-1000 years ago. During the years 1960--1980, the star 
was practically the spectroscopic twin of $\rho$\,Cas, because it displayed
the same emission and absorption features, varying with the pulsation phase.
It also underwent similar outbursts with the appearance of CO emission, which 
turned into absorption and then disappeared with the expansion and dilution of 
the ejected material \citep{2006ApJ...651.1130G}. However, in contrast to 
$\rho$\,Cas, from 1980 on V509\,Cas started to gradually heat up until about 2001
(Fig.\,\ref{fig2}). Concerning the structure of the small-scale environment, 
not much is currently known. 
The forbidden lines in our spectra (Fig.\,\ref{fig1}) show clearly 
double-peaked profiles for the [Ca\,{\sc ii}] lines, while the [O\,{\sc i}] 
lines appear single peaked at our spectral resolution. This behaviour is also 
seen in the B[e]SGs \citep[][Maravelias et al. this 
volume]{2010A&A...517A..30K, 2012MNRAS.423..284A} and might be assigned to a
(Keplerian) rotating disc or rings, in which the [Ca\,{\sc ii}] lines are 
formed closer to the star than the [O\,{\sc i}] $\lambda$6300 line. If true, 
V509\,Cas would be the second YHG with clear indication for an inner disk.

\section{Discussion and Conclusions}\label{discussion}

The fact that the hot YHGs in our sample seem to have disk-like structures, 
whereas the cool ones lack such evidence, might indicate a drastic change in
the mass-loss behaviour during the passage of the star through the Yellow Void
from spherical to axisymmetric mass-loss. As pulsations are believed to be the 
main mechanism responsible for enhanced mass loss and eruptions in YHGs 
\citep[see the overview by][]{1998A&ARv...8..145D}, this would require a change 
in the pulsation habit of the stars, which might be checked based on long-term 
observing campaigns.

During the past few decades, V1302\,Aql \citep{2002ARep...46..139K} and V509\,Cas 
\citep{2012A&A...546A.105N} displayed a considerable increase in their 
effective temperature, which seems to have slown down or even stopped.
\citet{2016MNRAS.459.4183K} reported that the kinematic picture and effective 
temperature of V1302\,Aql have been stable during 2001--2014. Our spectra from 
2015 show that the effective temperature of V509\,Cas has remained at the same 
level as in the beginning of 2001 (Fig.\,\ref{fig2}). These findings may
indicate that these objects are approaching the high-temperature boundary of 
the "first" instability region in the Yellow Void \citep{2012A&A...546A.105N}.
Such a conclusion is in line with the current position of the 
stars in the HRD (Fig.\,\ref{fig3}).

The presence of discs around these hot YHGs has further implications.
As these objects will continue to lose mass during their passage through the 
second part of the Yellow Void instability domain, the accumulation of material 
in the equatorial plane might continue. As soon as the star will reach the blue,
stable edge of the Yellow Void, it could have deposited enough material into 
the disk (or rings) to appear as a B[e]SG. Such an evolutionary link was 
already suggested for V1302\,Aql by \citet{Zickgraf1998} and 
\citet{2007ApJ...671.2059D}. But in light of the new proposition of a Keplerian
disk or rings around V509\,Cas based on the profile shapes of its forbidden
emission lines, it seems to be a valid suggestion even for a larger sample of hot YHGs.   

\acknowledgements A.A. and I.K. acknowledge financial support from the Estonian 
grant IUT40-1; M.K. and G.M. from GA\,\v{C}R (14-21373S) and  RVO:67985815.


\begin{thebibliography}{}
\expandafter\ifx\csname natexlab\endcsname\relax\def\natexlab#1{#1}\fi
\expandafter\ifx\csname url\endcsname\relax
  \def\url#1{\texttt{#1}}\fi
\expandafter\ifx\csname urlprefix\endcsname\relax\def\urlprefix{URL }\fi
\providecommand{\eprint}[2][]{\url{#2}}

\bibitem[{{Aret} et~al.(2016){Aret}, {Kraus}, \& {{\v
  S}lechta}}]{2016MNRAS.456.1424A}
{Aret}, A., {Kraus}, M., \& {{\v S}lechta}, M. 2016, \mnras, 456, 1424

\bibitem[{{Aret} et~al.(2012)}]{2012MNRAS.423..284A}
{Aret}, A., et~al. 2012, \mnras, 423, 284

\bibitem[{{Castro-Carrizo} et~al.(2007)}]{2007A&A...465..457C}
{Castro-Carrizo}, A., et~al. 2007, \aap, 465, 457

\bibitem[{{Davies} et~al.(2007){Davies}, {Oudmaijer}, \&
  {Sahu}}]{2007ApJ...671.2059D}
{Davies}, B., {Oudmaijer}, R.~D., \& {Sahu}, K.~C. 2007, \apj, 671, 2059

\bibitem[{{de Jager}(1998)}]{1998A&ARv...8..145D}
{de Jager}, C. 1998, \aapr, 8, 145

\bibitem[{{de Jager} \& {Nieuwenhuijzen}(1997)}]{1997MNRAS.290L..50D}
{de Jager}, C., \& {Nieuwenhuijzen}, H. 1997, \mnras, 290, L50

\bibitem[{{Ekstr{\"o}m} et~al.(2012)}]{2012A&A...537A.146E}
{Ekstr{\"o}m}, S., et~al. 2012, \aap, 537, A146

\bibitem[{{Gorlova} et~al.(2006)}]{2006ApJ...651.1130G}
{Gorlova}, N., et~al. 2006, \apj, 651, 1130

\bibitem[{{Klochkova} et~al.(2002)}]{2002ARep...46..139K}
{Klochkova}, V.~G., et~al. 2002, Astronomy Reports, 46, 139

\bibitem[{{Klochkova} et~al.(2016)}]{2016MNRAS.459.4183K}
--- 2016, \mnras, 459, 4183

\bibitem[{{Kraus} et~al.(2007){Kraus}, {Borges Fernandes}, \& {de
  Ara{\'u}jo}}]{2007A&A...463..627K}
{Kraus}, M., {Borges Fernandes}, M., \& {de Ara{\'u}jo}, F.~X. 2007, \aap, 463,
  627

\bibitem[{{Kraus} et~al.(2010){Kraus}, {Borges Fernandes}, \& {de
  Ara{\'u}jo}}]{2010A&A...517A..30K}
--- 2010, \aap, 517, A30

\bibitem[{{Kraus} et~al.(2016)}]{2016A&A...Kraus}
{Kraus}, M., et~al. 2016, \aap, forthcoming

\bibitem[{{Lobel} et~al.(2003)}]{2003ApJ...583..923L}
{Lobel}, A., et~al. 2003, \apj, 583, 923

\bibitem[{{McGregor} et~al.(1988){McGregor}, {Hyland}, \&
  {Hillier}}]{1988ApJ...334..639M}
{McGregor}, P.~J., {Hyland}, A.~R., \& {Hillier}, D.~J. 1988, \apj, 334, 639

\bibitem[{{Nieuwenhuijzen} et~al.(2012)}]{2012A&A...546A.105N}
{Nieuwenhuijzen}, H., et~al. 2012, \aap, 546, A105

\bibitem[{{Oksala} et~al.(2013)}]{2013A&A...558A..17O}
{Oksala}, M.~E., et~al. 2013, \aap, 558, A17

\bibitem[{{Oudmaijer} et~al.(1996)}]{1996MNRAS.280.1062O}
{Oudmaijer}, R.~D., et~al. 1996, \mnras, 280, 1062

\bibitem[{{Oudmaijer} et~al.(2009)}]{2009ASPC..412...17O}
--- 2009, in The Biggest, Baddest, Coolest Stars, edited by D.~G.
  {Luttermoser}, B.~J. {Smith}, \& R.~E. {Stencel}, vol. 412 of ASP Conference
  Series, 17

\bibitem[{{Schuster} et~al.(2006){Schuster}, {Humphreys}, \&
  {Marengo}}]{2006AJ....131..603S}
{Schuster}, M.~T., {Humphreys}, R.~M., \& {Marengo}, M. 2006, \aj, 131, 603

\bibitem[{{\v{S}lechta} \& {\v{S}koda}(2002)}]{2002PAICz..90....1S}
{\v{S}lechta}, M., \& {\v{S}koda}, P. 2002, Publ. Astron. Inst. Acad. Sci.
  Czech Republic, 90, 1

\bibitem[{{Tiffany} et~al.(2010)}]{2010AJ....140..339T}
{Tiffany}, C., et~al. 2010, \aj, 140, 339

\bibitem[{{Zickgraf}(1998)}]{Zickgraf1998}
{Zickgraf}, F.-J. 1998, {H}abilitation thesis,
  {R}uprecht-{K}arls-{U}niversit{\"{a}}t, {H}eidelberg

\bibitem[{{Zickgraf} et~al.(1986)}]{1986A&A...163..119Z}
{Zickgraf}, F.-J., et~al. 1986, \aap, 163, 119

\end{thebibliography}
\end{document}